# Decay Rates of Metastable States in Cubic Potential by Variational Perturbation Theory

H. Kleinert        I. Mustapic

Institut für Theoretische Physik
Freie Universität Berlin
Arnimallee 14     D-14195 Berlin

February 22, 1995

## Abstract

Variational perturbation theory is used to determine the decay rates of metastable states across a cubic barrier of arbitrary height. For high barriers, a variational resummation procedure is applied to the complex energy eigenvalues obtained from a WKB expansion; for low barriers, the variational resummation procedure converts the non-Borel-summable Rayleigh-Schrödinger expansion into an exponentially fast convergent one. The results in the two regimes match and yield very accurate imaginary parts of the energy eigenvalues. This is demonstrated by comparison with the complex eigenvalues from solutions of the Schrödinger equation via the complex-coordinate rotation method.

# Introduction

A particle in a one-dimensional anharmonic oscillator with a cubic potential is a popular model for tunneling processes with negligible backscattering. It has been widely used in the calculation of non-zero temperature quantum transition rates in the presence of friction [1, 2]. All such studies have been restricted to the regime of high barriers, where saddle-point and instanton methods are applicable. A typical expansion around an infinitely high barrier for a classical transition rate in a dissipative medium is given in [3].

In the last year, the situation has greatly been improved with the development of a variational perturbation expansion (VPE) for tunneling rates [4, 5]. On the one hand, the available tunneling calculations at high barriers were extended to lower barriers. On the other hand, the regime of very low barriers became accessible, where metastable states become short-lived and the decay proceeds by *sliding* rather than tunneling. The results converge rapidly for all barrier heights, as has been demonstrated for the anharmonic oscillator with an $x^4$-potential [5, 6]

The purpose of the present paper is to apply the variational method to the cubic potential and to give imaginary parts for the energy eigenvalues at *all* barrier heights, i.e., for all values of the coupling constants $\lambda$. In this way we go beyond earlier results by Drummond [7] who applied Borel-like methods to high-order expansions and was restricted to the regime of intermediate barrier heights.

Our inputs for the variational treatment are: a WKB-expansion for a high barrier and the Rayleigh-Schrödinger perturbation series when the barrier is low.

We have reformulated the WKB-method in a way that makes it accessible to computer algebra. This permitted us to determine the expansion coefficients around the WKB results up to $7^{\text{th}}$ order in $\lambda^2$. These coefficients by themselves do not help to improve the results since the semiclassical expansion is divergent. The variational approach resums the expansion and leads, even at the level of a first-order WKB-correction, to a remarkable improvement of the results.

In the low-barrier sliding regime, i.e., for not too small values of the coupling constant $\lambda$, a first-order VPE yields the imaginary parts of all eigenvalues $E_n(\lambda)$ within $\sim 10\%$ of the exact result (see Figure 1). This is all the more remarkable as no first order in $\lambda$ appears in the bare perturbation series which contains only even orders in $\lambda$, for symmetry reasons. To seventh order in $\lambda$, the results deviate by $\sim 5\times 10^{-5}$ (see Table 1). In the limit of infinitely large $\lambda$, a simple scaling argument shows that the eigenvalues grow like $\kappa\,\lambda^{2/5}$. The VPE reproduces this behavior and yields excellent results for $\kappa$ (see Table 2).

Let us remark that the cubic interaction has also attracted a certain amount of interest in quantum field theory, since a $\phi^3$-theory in six dimensions shares certain properties with quantum chromodynamics such as renormalizability and asymptotic freedom [8]. In the field theoretical context, the instability of the potential due to tunneling is generally discarded by *defining* the theory to be identical to its perturbative expansion, where all results are real numbers. The present method should eventually be generalized to field theory and help defining the $\phi^3$-theory beyond perturbation theory.

# 1 General Description of Variational Perturbation Expansion

We first describe the construction of the VPE in the regime of low barriers. Recall standard perturbation theory, which is based on splitting $H$ in two parts:
$$H_0 + H_{\text{int}},$$
whose first term can be analyzed exactly whereas the second is treated as a small correction. For an anharmonic oscillator with a quartic potential, the conventional splitting into

$$\begin{aligned} H_0^{\text{conv}} &= -\frac{1}{2}\frac{d^2}{d\,x^2} + \frac{\omega^2}{2}\,x^2, \\ H_{\text{int}}^{\text{conv}} &= gx^4, \end{aligned} \quad (1)$$

yields, for each eigenvalue, an asymptotic series in $g$ which is numerically useless unless $g$ is extremely small. In the VPE the splitting is

$$\begin{aligned} H_0^{\text{VPE}} &= -\frac{1}{2}\frac{d^2}{d\,x^2} + \frac{\Omega^2}{2}\,x^2 \\ H_{\text{int}}^{\text{VPE}} &= \frac{\omega^2-\Omega^2}{2g}\,x^2 + x^4. \end{aligned} \quad (2)$$

The free part of the Hamiltonian $H_0$ contains a trial frequency to be optimized systematically. The perturbation expansion in powers of $H_{\text{int}}^{\text{VPE}}$ for the $n$th excited energy eigenvalues $E^{(n)}$ yields an $\Omega$-dependent expressions:

$$E^{(n)\text{VPE}} = E^{(n)}(g, \Omega, \omega, N), \qquad (3)$$

where $N$ indicates the order in $g$ up to which the expansion has been performed. Since the final result must be independent of $\Omega$, an optimal energy is found by choosing that value of $\Omega$ for which the result has the weakest $\Omega$-dependence.[1] This criterion can be implemented in a number of ways: Usually one simply looks for a local extremum of $E^{(n)}(g, \Omega, \omega, N)$, i.e., one solves

$$\frac{d}{d\Omega} E^{(n)}(g, \Omega, \omega, N) = 0 \qquad (4)$$

for $\Omega$ which defines $\Omega$ to be a function of $g$, $N$ and $n$. The values

$$E^{(n)}(g, \Omega(g, N, n), \omega, N)$$

are good approximations to the exact eigenvalues even for $N = 1$, in particular, the exact large-$g$ behavior is qualitatively reproduced ($E^{(n)} = \kappa\, g^{1/3}$, as $g \to \infty$), and the constant $\kappa$ is astonishingly well approximated ($\kappa_{k=1}^{\text{VPE}} = 0.6814$, compared with the exact value $0.667986$).

For the quartic anharmonic oscillator, the criterion (4) can be applied only to odd orders $N$. Otherwise no zeros are found. In this case, the minimal sensitivity on $\Omega$ is given by

$$\frac{d^2}{d\Omega^2} E^{(n)}(g, \Omega, \omega, N) = 0 \qquad (5)$$

For large $N$, both conditions have many solutions. Then an optimal approximation is obtained by searching for *plateaux* in $\Omega$, where the $\Omega$-dependence is very weak. These plateaux may contain very smooth local extrema [6]. If the extrema are plotted in the $N$-$\Omega$-plane, the lower envelope of all extrema gives frequencies with the fastest convergence (exponentially fast even for infinite coupling strength).

In this paper, a similar treatment will be given for the cubic anharmonic oscillator, where the asymmetry of the potential requires a few modifications.

---

[1]This has been called the *Principle of Minimal Sensitivity* [9]. There exists also a slightly different, closely related principle of *Fastest Apparent Convergence* [10].

In Section 2.2, the $\Omega$-dependent energies (3) are determined from the *conventional* perturbation coefficients by a simple extension of the reexpansion trick developed in [5].

A convergence proof for the VPE method applied to Rayleigh-Schrödinger (RS) perturbation expansions based on a dispersion relation is given in [11] (see also [12], where the convergence of the variationally resummed expansion for the partition function is proved). The observed convergence is, however, much better than the bound given in the proof (cf. [5]). It is exponentially fast even for infinitely strong couplings.

## 2 Cubic Oscillator

We now consider the anharmonic oscillator with a cubic potential and a Hamiltonian:
$$H = -\frac{1}{2}\frac{d^2}{dx^2} + \frac{\omega^2}{2}x^2 - \lambda x^3. \qquad (6)$$

We choose energy units such that $\omega = 1$. Due to the asymmetry of the potential, the Hamiltonian must be split in a more general way than in (2):
$$H = H_0 + \lambda V, \qquad (7)$$

with
$$\begin{aligned} H_0 &= -\frac{1}{2}\frac{d^2}{dx^2} + \frac{\Omega^2}{2}x^2 + \Omega^2 x_0 x \\ V &= -\frac{\Omega^2}{\lambda}x_0 x + \frac{1-\Omega^2}{2\lambda}x^2 + x^3. \end{aligned} \qquad (8)$$

A second variational parameter $x_0$ permits us to choose an optimal position of the potential minimum. The VPE-improved eigenvalues are evaluated by solving the following set of equations for $x_0$ and $\Omega$:
$$\begin{aligned} \frac{\partial}{\partial \Omega} E^{(n)}(\lambda, \Omega, x_0, N) &= 0 \\ \frac{\partial}{\partial x_0} E^{(n)}(\lambda, \Omega, x_0, N) &= 0. \end{aligned} \qquad (9)$$

and inserting the solutions into
$$E^{(n)\mathrm{VPE}}(\lambda) = E^{(n)}(\lambda, \Omega(\lambda, N, n), x_0(\lambda, N, n), N).$$

Excellent approximations to the complex eigenvalues of the system are obtained by choosing the proper complex zeros of Eqs. (9). The functions $x_0(\lambda)$ and $1 - \Omega^2$ turn out to be functions of order $\mathcal{O}(\lambda)$ and $\mathcal{O}(\lambda^2)$, respectively so that $V$ can be regarded as being of order $\lambda^0$.

As a motivation to this splitting we recall that the VPE can be extended from energy eigenvalues to path integrals [5, 13, 14], where the lowest approximation reduces to the Feynman-Kleinert variational approach (FKVA) [15, 16] which is a powerful tool for the approximate calculation of partition functions, particle distributions, etc. The path integral in that approach contains an integration parameter, the path average

$$x_0 = \frac{1}{\beta} \int_0^\beta x(\tau) \, d\tau,$$

to be integrated over at the end. The partition function of a quantum mechanical particle

$$Z = \int \mathcal{D}x \, \exp\left[-\int_0^\beta \left\{\frac{\dot{x}(\tau)^2}{2} + V(x(\tau))\right\} d\tau\right] \tag{10}$$

is rewritten as an integral over $x_0$:

$$Z = \int_{-\infty}^{\infty} \frac{e^{-\beta V_{\text{eff}}(x_0)}}{\sqrt{2\pi\beta}} \, dx_0. \tag{11}$$

with the *effective classical potential* $V_{\text{eff}}(x_0)$ defined by the restricted path integral

$$\frac{1}{\sqrt{2\pi\beta}} e^{-\beta V_{\text{eff}}(x_0)} = \int \mathcal{D}x \, \delta\left(x_0 - \frac{1}{\beta}\int_0^\beta x(\tau)\, d\tau\right) \times$$
$$\exp\left[-\int_0^\beta \left\{\frac{\dot{x}(\tau)^2}{2} + \frac{\Omega^2}{2}(x(\tau) - x_0)^2\right\} d\tau\right] \times$$
$$\exp\left[-\int_0^\beta \left\{-\frac{\Omega^2}{2}(x(\tau) - x_0)^2 + V[x(\tau)]\right\} d\tau\right]. \tag{12}$$

The second exponential is treated as a perturbation, $\Omega$ is optimized and a highly improved expansion is obtained.

The *ground state energy* $E^{(0)}$ of the particle can be calculated from $Z$ via the relation

$$E^{(0)} = -\lim_{\beta \to \infty} \frac{\ln Z}{\beta}. \tag{13}$$

But in this limit, the integral (11) can be evaluated in the saddle-point-approximation, which means $E^{(0)}$ is simply equal to the minimum of $V_{\text{eff}}(x_0)$. Now, $V_{\text{eff}}(x_0)$ differs from the classical potential in that it contains information about quantum mechanical fluctuations, which tend to smear out the classical potential. This means barrier heights in the effective potential are lower than in its classical counterpart[2]. If the classical barrier is low enough, it will be even completely removed by the fluctuations, and there is our point: As the coupling constant $\lambda$ of the cubic oscillator is increased, the barrier height decreases and at a certain point, the minimum of the well will disappear from the effective potential. Then, the corresponding saddle-point in (11) will move into the complex plane and we shall obtain the desired complex eigenvalue.

Obviously, the effective classical potential of the cubic oscillator can be found from a variational approach only if the initial harmonic oscillator Hamiltonian $H_0$ has, in addition to the trial frequency $\Omega$, a variable bottom position as in (8).

This argument can be generalized to include excited states [13].

## 3 The Sliding Regime

### 3.1 The First Order in $\lambda$

We perform the Rayleigh-Schrödinger expansion of (7) to seventh order in $\lambda$. For greater transparency, we give first the lowest-order results. The associated VPE equations are then particularly simple and the optimal values of $\Omega$ and $x_0$ are good initial values when searching for extrema at higher orders.

The first order approximation to $E^{(n)}$ is:

$$E^{(n)}(\lambda, \Omega, x_0, 1) = (n + 1/2)\left(\frac{1 + \Omega^2}{2\Omega} - \frac{3\lambda x_0}{\Omega}\right) + \frac{x_0^2}{2} - \lambda x_0^3. \tag{14}$$

---

[2]In the case of the hydrogen atom, $V_{eff}$ loses the $1/r$-singularity. This is an appealing way to explain from a new perspective the stability of matter ([17], ch. 12).

The optimization conditions (9) amount to

$$\Omega^5 - \Omega + 36(n+1/2)\lambda^2 = 0,$$
$$\frac{1 - \sqrt{1 - 36\lambda^2(n+1/2)/\Omega}}{6\lambda} = x_0. \quad (15)$$

if we use as a supplementary condition that $x_0$ should be zero when $\lambda = 0$. The proper solution of Eq. (15) is found by following the evolution of the zero which is closest to $\Omega = 1$ for small $\lambda$. For $n = 0$ all results are real as long as $\lambda < 0.18$ beyond which value it acquires an imaginary part. This is due to the fact that the imaginary part of the energy — which can be interpreted as a tunneling rate — vanishes exponentially for $\lambda \to 0$. Obviously, Eqs. (15), being polynomial in nature, cannot reproduce exponentially small terms. Instead they drop to zero. We can give a physical interpretation of this behavior. The VPE is not capable of describing genuine tunneling, whose rate has the behavior $e^{-C/|\lambda|^2}$ and possesses no power series expansion. It can only account for the process of *sliding* down into the potential abyss, which sets in when the real part of the eigenvalue under consideration lies above the potential barrier.

Figure 1 shows the ground-state results for perturbation orders $N = 1, 3, 5, 7$ at the sliding-tunneling-divide. Just as in the case of the quartic perturbation, Eqs. (15) do not yield useful results for even $N$. The higher order results do not necessarily drop to zero for small $\lambda$, but cease to converge to the correct result beyond the sliding regime. In accordance with this, Fig. 2 shows that our approximation yields good approximations closer to $\lambda = 0$ as the quantum number rises.

## 3.2 Higher Orders

On dimensional grounds, the conventional Rayleigh-Schrödinger series for the $n$th excited state is an expansion in powers of $\lambda/\omega^{5/2}$:

$$E^{(n)}(\lambda) = \omega \sum_{j=0}^{\infty} E_j^{(n)} \left(\frac{\lambda}{\omega^{5/2}}\right)^j. \quad (16)$$

All odd coefficients vanish. The coefficients are most easily calculated via the following recursion relation [7, 18]:

$$(j - m) B_{m,j}^{(n)} = \frac{1}{2} (j+1)(j+2) B_{m,j+2}^{(n)} + B_{m-1,j-3}^{(n)} - \sum_{k=1}^{m-1} B_{k,2}^{(n)} B_{m-k,j}^{(n)}$$
$$E_j^{(n)} = -B_{j,2}^{(n)} \tag{17}$$

for states with even quantum number $n$, and

$$(j - m) B_{m,j}^{(n)} = \frac{1}{2} (j+1)(j+2) B_{m,j+2}^{(n)} + B_{m-1,j-3}^{(n)} - 3 \sum_{k=1}^{m-1} B_{k,3}^{(n)} B_{m-k,j}^{(n)}$$
$$E_j^{(n)} = -3 B_{j,3}^{(n)} \tag{18}$$

for odd $n$.

It is easy to show that the reorganized series corresponding to the splitting (8) is generated from such a series by the following operation generalizing the trick in Ref. [5]:

$$e^{\lambda a \, \partial/\partial \Omega^2} e^{-\lambda b \, \partial/\partial j} \left[ -j \, \xi + \frac{\Omega^2}{2} \xi^2 - \lambda \, \xi^3 + \right.$$
$$\left. \Omega \left(1 - \frac{12 \lambda j}{\Omega^4}\right)^{1/4} \sum_{\nu=0}^{\infty} E_{2\nu}^{(n)} \left( \frac{\lambda^2}{\Omega^5 (1 - 12 j \lambda/\Omega^4)^{5/4}} \right)^{\nu} \right]. \tag{19}$$

with

$$\xi := \frac{\Omega^2 - \sqrt{\Omega^4 - 12 j \lambda}}{6 \lambda},$$
$$j := \Omega^2 x_0. \tag{20}$$

Upon expansion to the desired order in $\lambda$, $a$ and $b$ have to be replaced by:

$$a \to \frac{1 - \Omega^2}{\lambda},$$
$$b \to \frac{j}{\lambda}.$$

This shows that the only input needed for a VPE calculation are the coefficients of the divergent series (16). The VPE performs a resummation of this series which turns out to be convergent.

# 4  The Tunneling Regime

For large barriers, i.e., for small $\lambda$, the imaginary part of the energy eigenvalues is found as usual from the WKB-wave-functions of the cubic oscillator, matched asymptotically to harmonic oscillator wave functions around the origin. For the cubic potential

$$V(x) = \frac{\omega^2}{2} x^2 - \lambda\, x^3,$$

this well-known method [19] leads to the following result:

$$\operatorname{Im} E^{(n)}(\lambda) =: \epsilon_{\mathrm{WKB}}(\lambda, n) = \frac{8^n \omega}{\sqrt{\pi}\, n!} \left(\frac{\omega^5}{\lambda^2}\right)^{n+1/2} \exp\left[-\frac{2\omega^5}{15\lambda^2}\right], \qquad (21)$$

which is correct to zeroth order in $\lambda$. For the inverted quartic oscillator, the WKB-calculation has been pushed to first order in the coupling constant by Bender and Wu [18]. Inspired by the method they present in the Appendix of [18], we have carried the WKB-calculation for the cubic oscillator to seventh order in $\lambda^2$ using computer algebra (see Appendix A where we described our method which is more efficient than the original version by Bender and Wu). To first order, we obtain (see Appendix A.2 for the seventh order)

$$\operatorname{Im} E^{(n)}(\lambda) = \epsilon_{\mathrm{WKB}}(\lambda, n) \left\{ 1 - \left(\frac{169}{16} - \frac{201}{8} n - \frac{141}{8} n^2\right) \frac{\lambda^2}{\omega^5} + \mathcal{O}[\lambda^4] \right\} \quad (22)$$

In Appendix B we have derived the first-order correction factor within the path integral formalism [20].

The result is now improved by an extension of the variational method developed in Refs. [4, 5]. The logarithm of the imaginary part (22) is written down for the Hamiltonian $H_0$ and evaluated at the optimal values of the variational parameters $\Omega$ and $j$ obtained by the procedure in the last section. The VPE equations (9) yield improved values for the imaginary parts of the eigenvalues in the tunneling regime [21]. Appendix B shows that this procedure is legitimate for the ground-state. For the excited states it turns out to match reasonably well onto the results from the preceding section and is validated by success. It also compares very favorably with results for the cubic oscillator obtained by more conventional means to which we now turn our attention:

| $\lambda$ | $E^{(0)\,\mathrm{CCR}}$ | $E^{(0)\,\mathrm{VPE}}$ |
|---:|---|---|
| 0.18 | $0.43386176 + 0.02524252\,i$ | $0.43373 + 0.02530\,i$ |
| 1 | $0.61288846 + 0.40859267\,i$ | $0.61285 + 0.40861\,i$ |
| 100 | $3.89396500 + 2.82900663\,i$ | $3.89383 + 2.82916\,i$ |

| $\lambda$ | $E^{(4)\,\mathrm{CCR}}$ | $E^{(4)\,\mathrm{VPE}}$ |
|---:|---|---|
| 0.07 | $3.87967181 + 0.05672784\,i$ | $3.87705 + 0.05811\,i$ |
| 0.1 | $3.59046484 + 0.87272058\,i$ | $3.59583 + 0.87116\,i$ |
| 1 | $8.13685894 + 5.82497954\,i$ | $8.13549 + 5.8239\,i$ |
| 10 | $20.4997005 + 14.8887626\,i$ | $20.4963 + 14.8863\,i$ |
| 20 | $27.051325 + 19.651713\,i$ | $27.0469 + 19.6485\,i$ |
| 50 | $39.03 + 28.35\,i$ | $39.0215 + 28.3501\,i$ |

Table 1: Comparison of VPE with the results of a numerical solution of the Schrödinger equation via the complex coordinate rotation method CCR for the quantum numbers $n = 0$ and $n = 4$. The VPE input was the seventh-order Rayleigh-Schrödinger expansion.

## 5   Comparison with Schrödinger Theory

A surprising feature of the VPE for complex eigenvalues is that no need arises to discuss boundary conditions. This is in contrast to the numerical solution of the Schrödinger equation, where a non-Hermitian radiation condition has to be imposed. We solved this equation using the complex coordinate rotation (CCR) method [22], applying one of the conventional techniques. We have chosen to use the elegant formalism of continued matrix fractions [23, 1]. Figure 2 shows the result of a VPE calculation based on a seventh-order perturbation expansion in $\lambda$ of (19) for the sliding regime, matched to a third-order expansion for the tunneling regime. The systematically improved WKB results from Appendix A.2 have not been plotted since they almost coincide with the variationally improved ones up to the sliding regime, where they become inapplicable. Table (1) shows numbers for the seventh-order sliding results and their excellent agreement with the numerical Schrödinger calculations.

| $n$ | $\kappa$ |
|---|---|
| 0 | $0.617159 + 0.448390\,i$ |
| 1 | $2.245672 + 1.631577\,i$ |
| 2 | $4.036069 + 2.932440\,i$ |
| 3 | $6.038479 + 4.387212\,i$ |
| 4 | $8.160972 + 5.929293\,i$ |

Table 2: For large coupling $\lambda$, the $E^{(n)}(\lambda)$ go like $E^{(n)}(\lambda) \to \kappa\,\lambda^{2/5}$. The table shows the coefficients $\kappa$ for the lowest five states, obtained from an 11$^{\text{th}}$-order VPE calculation.

## 6 Conclusion

We have successfully applied the VPE to an asymmetric system with complex eigenvalues. As a resummation method, VPE is more powerful than traditional techniques such as Borel-resummation, Padé-approximants, etc., since

- it gives good results already at the lowest perturbation order ,
- it is valid for arbitrarily large values of the coupling constant.

The asymmetry of the cubic potential produces a shift of the minimum of the effective classical potential as the coupling constant is increased. The same situation will be encountered in field theories with symmetry breaking.

A convergence-proof and an estimate of the remainder term are still missing. Both will probably have to rely on the dispersion relation [5, 10]:

$$\operatorname{Re} E^{(n)}(\lambda) = \omega\left(n + \frac{1}{2}\right) + \frac{\lambda}{\pi}\,\mathrm{P}\int_{-\infty}^{\infty} \operatorname{Im} E^{(n)}(x)\,\frac{1}{x(x-\lambda)}. \qquad (23)$$

A method to reproduce the results from appendix A on a path integral level, thus systematically improving on the harmonically matched semiclassical path integral calculation presented in [24, 25], remains still to be found. This is needed to extend the WKB results to quantum field theory.

It may prove an instructive intermediate step to try and find a recursion relation for the coefficients of the WKB-series which is similarly simple as the one given by Zinn-Justin in [26] for the systematic corrections to the semiclassical energy-level splitting in the double well potential.

An immediate application of the present results will be a calculation of the quantum mechanical transition rates at non-zero temperature. One can exploit the fact that the VPE readily yields complex energy eigenvalues for arbitrarily large quantum numbers. These can be used to find statistical averages of decay rates. We believe that this may ultimately pave the way to a similarly efficient variational method for the computation of *dissipative* quantum transition rates.

# A  Appendix: WKB-Calculation

## A.1  The Procedure

The small-$\lambda$-limit for the eigenvalues of the Schrödinger equation
$$\left[ -\frac{1}{2}\frac{d^2}{dx^2} + \frac{\omega^2}{2}x^2 - \lambda x^3 - E \right] \psi(x) = 0 \tag{24}$$
is related to the $\hbar \to 0$-limit of the semiclassical approximation by the rescaling
$$x \to x/\lambda.$$
The Schrödinger equation for the cubic oscillator becomes:
$$\left[ -\frac{1}{2}\frac{d^2}{dx^2} + \frac{(\omega^2/2)x^2 - x^3 - \lambda^2 E}{\lambda^4} \right] \psi(x) = 0. \tag{25}$$

The asymptotically correct behavior of the eigenvalues as $\lambda \to 0$ is obtained if one notes that the two turning points around the minimum coalesce in this limit, which means that, contrary to textbook-WKB, we must not approximate the potential linearly in the neighborhood of these turning points but have to approximate it quadratically. From a technical point of view, this facilitates our task since it means that, in *region A* of Fig. 3 we can expand our solution around harmonic oscillator wave functions instead of Airy-functions. In other words: In *region A*, the wave functions is sufficiently well approximated by the Rayleigh-Schrödinger wave function calculated to the desired order in $\lambda$.

For clarity's sake, we shall present our procedure for the first order in $\lambda^2$. The result for the seventh order can be found at the end of this appendix:
$$\psi_{\rm rs}(x) = \psi_0(x) + \lambda\,\delta\psi_1(x) + \lambda^2\,\delta\psi_2(x) + \mathcal{O}[\lambda^3].$$

The fact that $E$ scales with $\lambda^2$ is advantageous in another respect as well: In *region B* we have to calculate the WKB-solution of Eq. (24) to first WKB-order. Its exponentially decreasing part is:

$$\psi_{\text{WKB}}(x) = p(x)^{-1/2} \exp\left[-\int^x \left(p(\xi) - \frac{1}{2} p(\xi)^{-1/2} \frac{d^2}{d\xi^2} p(\xi)^{-1/2}\right) d\xi\right], \quad (26)$$

with

$$p(x) = \sqrt{\omega^2 x^2 - 2\lambda x^3 - 2E}.$$

and $E$ has to be given correctly to first order in $\lambda^2$. From perturbation theory we obtain:

$$E^{(n)}(\lambda) = \omega\left(n + \frac{1}{2}\right) - \frac{\lambda^2}{8\omega^4}(11 + 30n + 30n^2) + \mathcal{O}[\lambda^4].$$

In this form, $\psi_{\text{WKB}}$ contains unappealing elliptic integrals. The expressions can be readily processed by computer algebra if the WKB-integrand is expanded to second order in $E$ before performing the integration. Our rescaling shows that this renders the WKB-function with sufficient accuracy. Following Bender and Wu [18] we shall neglect the exponentially increasing part of the WKB-wave-function at the eigenvalue. Furthermore, the integration constant for the integral in Eq. (26) is chosen in such a way that, beyond the far (linear) turning point, the integral becomes purely imaginary up to terms that vanish for $x \to \infty$. This solution is not a good approximation to the true wave function near the far (linear) turning point, but, at a sufficient distance, it is valid on both sides since it may be continued analytically from left to right along a semi-circle in the complex plane, thus keeping at bay the singularity at the turning point (see Fig. 3).

Away from the potential well and its coalescing turning points — i.e., for $x \sim k/\lambda$ see Fig. 3 — $\psi_{\text{rs}}(x)$ and $\psi_{\text{WKB}}(x)$ match to second order in $\lambda$:

$$\psi_{\text{rs}}(k/\lambda) = \mathcal{C}\, \psi_{\text{WKB}}(k/\lambda) + \mathcal{O}[\lambda^3], \quad (27)$$

where:

$$\begin{aligned}
\mathcal{C} &= e^{-\omega^5/(15\lambda^2)} \sqrt{\frac{8^n \omega^{5n+7/2}}{(\lambda^2)^{n+1/2} \sqrt{\pi}\, n!}} \times \\
&\quad \left[1 + \frac{\lambda^2}{\omega^5}\left(-\frac{449}{96} - \frac{1553}{144} n - \frac{341}{48} n^2 + \frac{41}{36} n^3\right)\right].
\end{aligned} \quad (28)$$

The constant $k$ appears only in the next order in $\lambda^2$. Knowing the constant $\mathcal{C}$ we are done. The imaginary part of the eigenvalue can be calculated from the quantum mechanical current:

$$\mathrm{Im}\, E = \lim_{x \to \infty} \frac{1}{4i} \left( \psi(x) \frac{d}{dx} \psi^*(x) - \psi^*(x) \frac{d}{dx} \psi(x) \right) \bigg/ \int_{-\infty}^{\infty} |\psi(\xi)|^2 d\xi\,, \qquad (29)$$

In the numerator of Eq. (29) we may use the WKB wave-function in the limit $x \to \infty$ where it becomes exact and in the numerator $\psi$ may be replaced by $\psi_{\mathrm{rs}}$, the error being exponentially small. Hence:

$$\mathrm{Im}\, E = \mathcal{C}^2 / <\psi_{\mathrm{rs}}|\psi_{\mathrm{rs}}> + \mathcal{O}[\lambda^4], \qquad (30)$$

and the result in Eq. (22) follows now to second order in $\lambda^2$, because:

$$<\psi_{\mathrm{rs}}|\psi_{\mathrm{rs}}> = 1 + \frac{\lambda^2}{72\,\omega^5}\left(87 + 256\,n + 246\,n^2 + 164\,n^3\right) + \mathcal{O}[\lambda^4].$$

## A.2 The Seventh-Order Result

For simplicity we set $\omega = 1$. The seventh-order systematic correction to $\epsilon_{\mathrm{WKB}}(\lambda, n)$ is then given by

$$\begin{aligned}\mathrm{Im}\, E^{(n)}(\lambda) &= \epsilon_{\mathrm{WKB}}(\lambda, n)(1 + k_1\,\lambda^2 + k_2\,\lambda^4 + k_3\,\lambda^6 + \\ & \quad + k_4\,\lambda^8 + k_5\,\lambda^{10} + k_6\,\lambda^{12} + k_7\,\lambda^{14} + \mathcal{O}[\lambda^{16}]),\end{aligned}$$

where

$$\epsilon_{\mathrm{WKB}}(\lambda, n) = \frac{8^n}{\sqrt{\pi}\,n!}\left(\frac{1}{\lambda^2}\right)^{n+1/2} \exp\left[-\frac{2}{15\lambda^2}\right]$$

and

$$k_1 = -\frac{169}{16} - \frac{201}{8}n - \frac{141}{8}n^2$$

$$k_2 = -\frac{44507}{512} - \frac{23639}{128}n - \frac{81}{4}n^2 + \frac{12907}{64}n^3 + \frac{19881}{128}n^4$$

$$k_3 = -\frac{86071851}{40960} - \frac{24682861}{4096}n - \frac{20228607}{4096}n^2 + \frac{287481}{256}n^3 +$$

$$+\frac{5847253}{2048}n^4 + \frac{356307}{1024}n^5 - \frac{934407}{1024}n^6$$

$$k_4 = -\frac{189244716209}{2621440} - \frac{83086372869}{327680}n - \frac{52650975367}{163840}n^2 -$$

$$-\frac{4279463287}{32768}n^3 + \frac{4119876039}{65536}n^4 + \frac{842550663}{16384}n^5$$

$$-\frac{69299813}{8192}n^6 - \frac{119027547}{8192}n^7 + \frac{131751387}{32768}n^8$$

$$k_5 = -\frac{128830328039451}{41943040} - \frac{261646543133639}{20971520}n - \frac{423592869141879}{20971520}n^2 -$$

$$-\frac{19715332860521}{1310720}n^3 - \frac{15339960598811}{5242880}n^4 + \frac{1401355481021}{524288}n^5 +$$

$$+\frac{2560290063173}{2621440}n^6 - \frac{25460831311}{65536}n^7 - \frac{122374368543}{524288}n^8 +$$

$$+\frac{31204521765}{262144}n^9 - \frac{18576945567}{1310720}n^{10}$$

$$k_6 = -\frac{1027625748709963623}{6710886400} - \frac{232890795508816773}{335544320}n -$$

$$-\frac{111071124048881887}{83886080}n^2 - \frac{664185578830753997}{503316480}n^3 -$$

$$-\frac{214792314439921013}{335544320}n^4 - \frac{1384181879205177}{41943040}n^5 +$$

$$+\frac{6816589779335}{65536}n^6 + \frac{341591367423137}{20971520}n^7 -$$

$$-\frac{1077894391687337}{83886080}n^8 - \frac{47307658671163}{12582912}n^9 +$$

$$+\frac{16009695476127}{5242880}n^{10} - \frac{6433288475823}{10485760}n^{11} +$$

$$+\frac{873116441649}{20971520}n^{12}$$

$$k_7 = -\frac{933142404651555165943}{107374182400} - \frac{2302284992253612495297}{53687091200}n -$$

$$-\frac{34667142800220818412209}{375809638400}n^2 - \frac{55618044990614950369}{503316480}n^3 -$$

$$-\frac{81735871758067504249}{1073741824}n^4 - \frac{13377851713824147883}{536870912}n^5 +$$

$$+\frac{953124390791341475}{536870912}n^6 + \frac{64443774954315701}{16777216}n^7 +$$

$$+\frac{1044173780396492593}{9395240960}n^8 - \frac{791622589744759973}{2013265920}n^9 -$$

$$-\frac{7250316957748797}{134217728}n^{10} + \frac{1631225395815011}{20971520}n^{11} -$$

$$-\frac{6912202077410247}{335544320}n^{12} + \frac{397936750990707}{167772160}n^{13} -$$

$$-\frac{123109418272509}{1174405120}n^{14}$$

# B  Appendix: Path Integral Calculation of Corrections to Semiclassical Limit

The $n = 0$-result in Eq. (22) is reproduced to first order in $\lambda^2$ by expanding the exponential in (10), with the cubic oscillator potential inserted, around an instanton solution to second order in $\lambda$. Two correlation functions with respect to an instanton Green function (cf. [20]) have to be calculated. Using computer algebra one obtains up to terms which vanish exponentially as $\beta \to \infty$:

$$T_1 := \frac{1}{2}\int_{-\beta/2}^{\beta/2}dt\int_{-\beta/2}^{\beta/2}dt'\ <x(t)^3 x(t')^3> = \frac{11}{8}\beta - \frac{601}{336}$$

$$T_2 := -\frac{1}{\mathcal{A}_{cl}} \int_{-\beta/2}^{\beta/2} dt \int_{-\beta/2}^{\beta/2} dt' <x''_{cl}(t)\,x(t)\,x(t')^3> = -\frac{559}{56}. \qquad (31)$$

Part of the first value belongs to the real part of the partition function and has to be subtracted. This contribution is found by calculating the first correlation with the Green function corresponding to the trivial classical solution $x_{cl,tr}(t) \equiv 0$ which is:

$$G_{tr}(t,t') = \frac{1}{2\omega} e^{-\omega|t-t'|}. \qquad (32)$$

The result is:

$$T_3 := \frac{1}{2} \int_{-\beta/2}^{\beta/2} dt \int_{-\beta/2}^{\beta/2} dt' <x(t)^3\,x(t')^3>_{tr} = \frac{11}{8}\beta - \frac{29}{24}.$$

In the end we obtain:
$$T_1 + T_2 - T_3 = -\frac{169}{16}$$

calculating the classical action and the fluctuation determinant (see [5], ch. 17) one obtains to first order in $\lambda^2$:

$$\operatorname{Im} E^{(0)}(\lambda) = \frac{\omega}{\sqrt{\pi}} \sqrt{\frac{\omega^5}{\lambda^2}} \exp\left[-\frac{2\omega^5}{15\lambda^2}\right] \left(1 - \frac{169}{16}\lambda^2\right) + \mathcal{O}[\lambda^4] \qquad (33)$$

in agreement with (22) for $n = 0$.

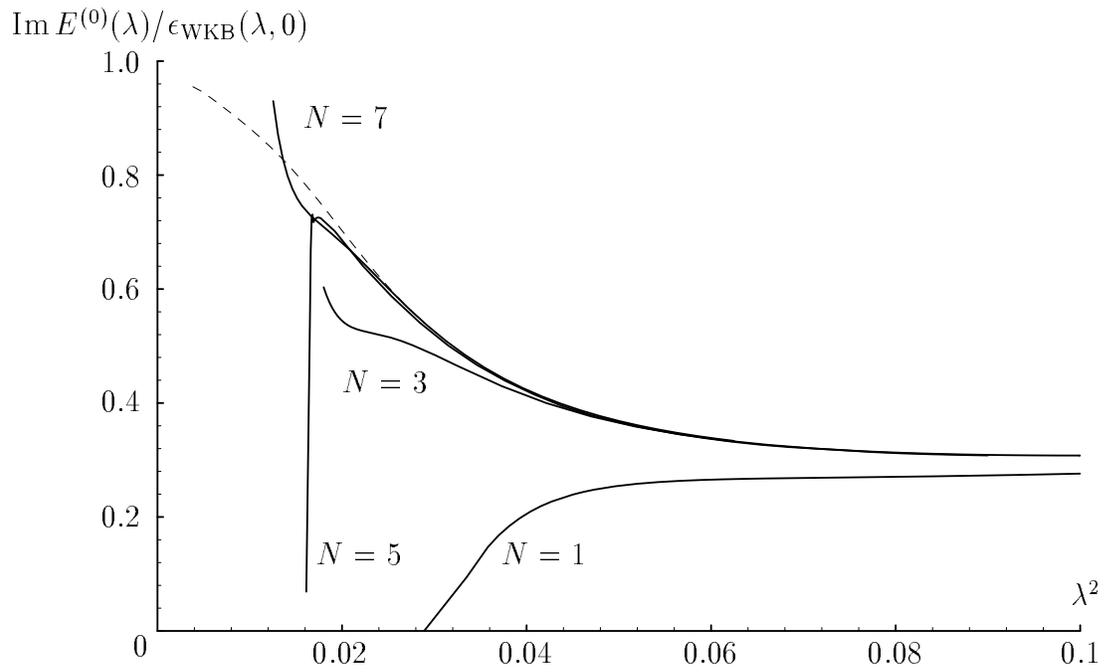

Figure 1: Convergence of the VPE for the imaginary part of the ground state energy around the tunneling-sliding divide. The dashed line is a numerical result obtained by complex coordinate rotation (see Section 4)

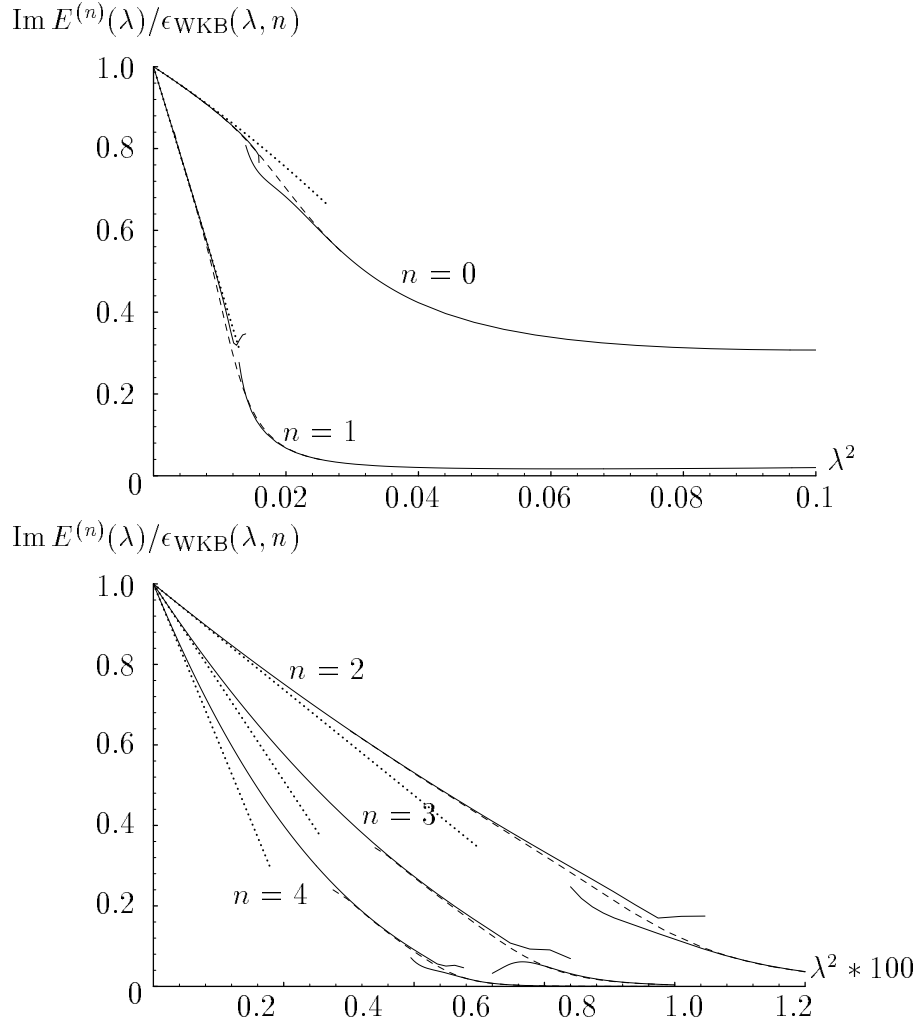

Figure 2: Imaginary parts of energy eigenvalues for quantum numbers $n = 0, 1, 2, 3, 4$ divided by zeroth-order WKB results. Solid lines on l.h.s.: variationally improved WKB results from a $3^{\text{rd}}$-order expansion in $\lambda$ ($5^{\text{th}}$ order for $n = 0$). On r.h.s.: improved RS results from a $7^{\text{th}}$-order expansion. Dashed lines: exact results from complex coordinate rotation. Dotted lines: $1^{\text{st}}$-order WKB result without variation ($2^{\text{nd}}$-order for $n = 0$).

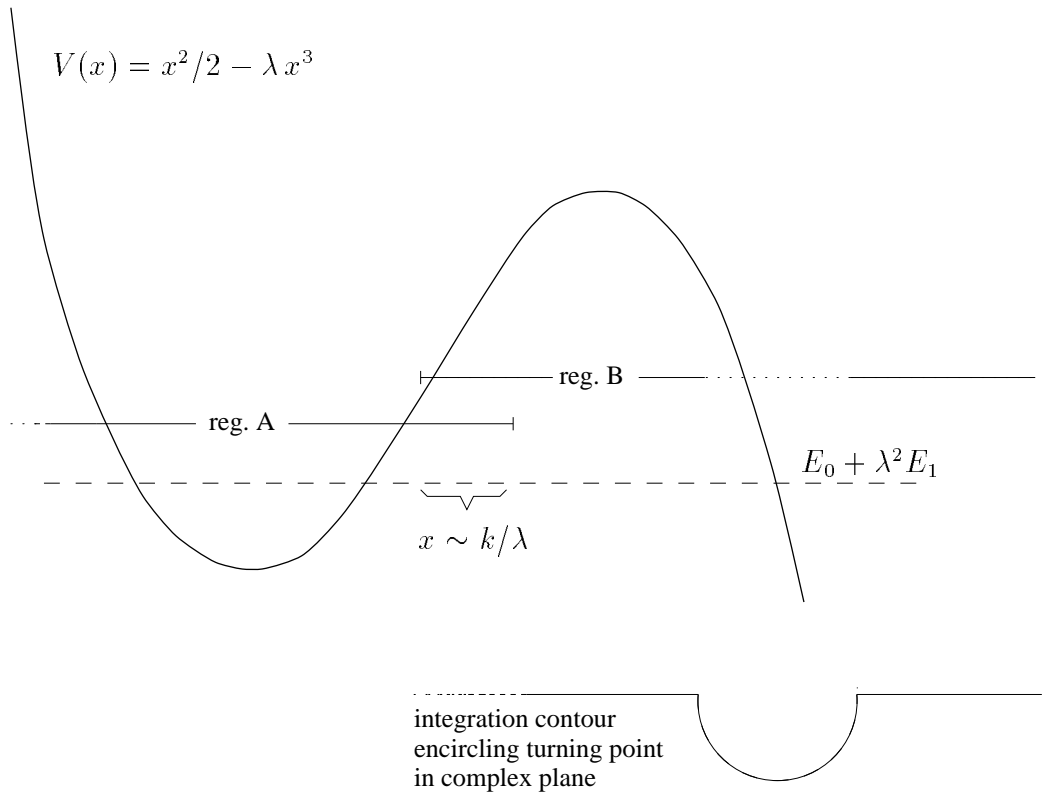

Figure 3: Tunneling Rate for High Barriers by WKB